%% file: TrueTransients.tex
\documentclass[twocolumn]{aastex631}

\usepackage{mathptmx}
\usepackage[T1]{fontenc}
\usepackage{ae,aecompl}
\usepackage{graphicx}	
\graphicspath{{.}} 
\usepackage{amsmath}	
\usepackage{amssymb}	
\usepackage{subfigmat}
\usepackage{enumerate}
\usepackage{multirow}
\usepackage{natbib}

\shorttitle{The JCMT Transient Survey: Faint Sources}
\shortauthors{D. Johnstone et al.}

\begin{document}

\title{The JCMT Transient Survey: Single Epoch Transients and Variability of Faint Sources}

\correspondingauthor{Doug Johnstone}
\email{Doug.Johnstone@nrc-cnrc.gc.ca}

\author[0000-0002-6773-459X]{Doug Johnstone}
\affiliation{NRC Herzberg Astronomy and Astrophysics, 5071 West Saanich Rd, Victoria, BC, V9E 2E7, Canada}
\affiliation{Department of Physics and Astronomy, University of Victoria, Victoria, BC, V8P 5C2, Canada}

\author[0000-0003-1618-2921]{Bhavana Lalchand}
\affiliation{Institute of Astronomy, National Central University, 300 Zhongda Road, Zhongli, Taoyuan 32001, Taiwan, R.O.C.}

\author[0000-0002-6956-0730]{Steve Mairs}
\affiliation{SOFIA Science Center, Universities Space Research Association, NASA Ames Research Center, Moffett Field, California 94035, USA}
\affiliation{East Asian Observatory, 600 N. A`oh\=ok\=u Place, Hilo, HI 96720, USA}

\author[0000-0001-8385-9838]{Hsien Shang}
\affiliation{Institute of Astronomy and Astrophysics, Academia Sinica, Taipei 10617, Taiwan, R.O.C.}

\author[0000-0003-0262-272X]{Wen Ping Chen}
\affiliation{Institute of Astronomy, National Central University, 300 Zhongda Road, Zhongli, Taoyuan 32001, Taiwan, R.O.C.}
\affiliation{Department of Physics,  
National Central University, 300 Zhongda Road, Zhongli, Taoyuan 32001, Taiwan, R.O.C.}

\author[0000-0003-4056-9982]{Geoffrey C.\ Bower}
\affiliation{Institute of Astronomy and Astrophysics, Academia Sinica, 645 N. A'ohoku Place, Hilo, HI 96720, USA}

\author[0000-0002-7154-6065]{Gregory J.\ Herczeg}
\affiliation{Kavli Institute for Astronomy \& Astrophysics, Peking University, Yiheyuan Lu 5, Haidian Qu, 100871 Beijing, China}
\affiliation{Department of Astronomy, Peking University, Yiheyuan 5, Haidian Qu, 100871 Beijing, China}

\author[0000-0003-3119-2087]{Jeong-Eun Lee}
\affiliation{School of Space Research, Kyung Hee University, 1732, Deogyeong-daero, Giheung-gu, Yongin-si, Gyeonggi-do 17104, Republic of Korea}

\author[0000-0001-8694-4966]{Jan Forbrich}
\affiliation{Centre for Astrophysics Research, University of Hertfordshire, College Lane, Hatfield, AL10 9 AB, UK}

\author{Bo-Yan Chen}
\affiliation{Institute of Astronomy, National Tsing Hua University, Taiwan}
\affiliation{Institute of Astronomy and Astrophysics, Academia Sinica, 645 N. A'ohoku Place, Hilo, HI 96720, USA}

\author[0000-0003-1894-1880]{Carlos Contreras-Pe\~na}
\affiliation{Department of Physics and Astronomy, Seoul National University, 1 Gwanak-ro, Gwanak-gu, Seoul 08826, Republic of Korea}
\affiliation{School of Space Research, Kyung Hee University, 1732, Deogyeong-daero, Giheung-gu, Yongin-si, Gyeonggi-do 17104, Republic of Korea}

\author[0000-0001-6047-701X]{Yong-Hee Lee}
\affiliation{School of Space Research, Kyung Hee University, 1732, Deogyeong-daero, Giheung-gu, Yongin-si, Gyeonggi-do 17104, Republic of Korea}

\author{Wooseok Park}
\affiliation{School of Space Research, Kyung Hee University, 1732, Deogyeong-daero, Giheung-gu, Yongin-si, Gyeonggi-do 17104, Republic of Korea}

\author{Colton Broughton}
\affiliation{Department of Physics and Astronomy, University of Victoria, Victoria, BC, V8P 5C2, Canada}

\author{Spencer Plovie}
\affiliation{Department of Physics and Astronomy, University of Victoria, Victoria, BC, V8P 5C2, Canada}

\author{The JCMT Transient Team}


\begin{abstract}
Short-duration flares at millimeter wavelengths provide unique insights into the strongest magnetic reconnection events in stellar coronae, and combine with longer-term variability to introduce complications to next-generation cosmology surveys.  We analyze 5.5 years of JCMT Transient Survey 850\micron\ submillimeter monitoring observations toward eight Gould Belt star-forming regions to search for evidence of transient events or long-duration variability from faint sources. The eight regions (30\arcmin\ diameter fields), including $\sim 1200$ infrared-selected YSOs, have been observed on average 47 times with integrations of approximately half an hour, or one day total spread over 5.5 years. Within this large data set, only two robust faint source detections are recovered: JW\,566 in OMC\,2/3 and MGM12\,2864 in NGC\,2023. JW\,566, a Class\,II T\,Tauri binary system previously identified as an extraordinary submillimeter flare, remains unique, the only clear single-epoch transient detection in this sample with a flare eight times bright than our {$\sim 4.5$-sigma} detection threshold of 55\,mJy/beam. The lack of additional recovered flares intermediate between JW\,566 and our detection limit is puzzling, if smaller events are more common than larger events. In contrast, the other submillimeter variable identified in our analysis, Source\,2864, is highly variable on all observed timescales. Although Source\,2864 is occasionally classified as a YSO, the source is most likely a blazar.  The 
degree of variability across the electromagnetic spectrum may be used to aid source classification.
\end{abstract}


\section{Introduction}\label{sec:introduction}

Young stellar objects (YSOs) are known for their variability due to both accretion phenomena and magnetic activity.   Young low-mass pre-main sequence stars are fully convective and therefore develop magetospheres from stellar dynamos. High-energy events driven by the magnetosphere's magnetic field interacting with itself or its surroundings are expected to occur, producing powerful stellar flares. Monitoring the timescales and amplitude changes of their emission over time provides a useful probe of these physical processes.

Coronal flares are driven by magnetic reconnection of large loops of the field that protrude from the stellar surface. The magnetic reconnection converts magnetic energy into gas kinetic energy and bulk plasma motions. Most of the energy released is thermalized and radiated away as thermal emission, which can be measured using the soft x-ray emission from the corona or ultraviolet line emission from the chromosphere \citep{benz10}. A fraction of the released energy is emitted at radio frequencies as gyrosynchrotron radiation \citep{waterfall19}, manifesting as radio flares. An empirical scaling relation, $L_X/L_R \sim 10^{15\pm 1}$\,Hz links the two across several orders of magnitude, although the X-rays saturate at $\log L_X/L_{\rm bol} \sim -3$ \citep{gudel93}.  

Stellar flares are typically explained as magnetic reconnection in the stellar magnetosphere. However, for young stars the largest events, such as those equivalent to coronal mass ejections, may arise from enormous loops. These loops could potentially couple to the surrounding accretion disk, if present, and enhance the magnitude of the flare \citep{for16,for17}. Compared with main-sequence stars, flares have an elevated importance for T Tauri stars, with X-ray flare luminosities typically ranging from $L_X \sim 10^{28\mbox{--}31}$\,erg~s$^{-1}$, comparable to the solar maximum $L_X \sim 10^{27}$\,erg~s$^{-1}$ \citep{feigelson99}. More recently, \citet{Getman2021} identified a sample of 1000 super-flares with $L_X \sim 10^{30.5\mbox{--}34}$\,erg~s$^{-1}$ from YSOs, detected on both disk-bearing and diskless systems and across a wide-range of evolutionary stages, including protostars. Since many of these large flares are detected on diskless stars, the emission volume of the magnetic loops either does not require an anchor or may be amplified by anchoring the field to a nearby magnetized stellar or substellar companion \citep{lin22}.

Previous observations of radio flares from YSOs have been reported at mm \citep{bower2003,furuya2003,massi2006,salter2008} and cm wavelengths \citep{for08,for17}. The brightest of these flares reached $L_R \sim 10^{19}$\,erg/s/Hz \citep{bower2003}, many orders of magnitude higher than the M-type star radio flaring events monitored in the mm by \citet{Macgregor2018, Macgregor2020}, and comparable to the highest super-flare X-ray luminosities, assuming the empirical scaling-relation. With the advent of all-sky monitoring campaigns at mm wavelengths by the Atacama Cosmology Telescope (ACT) and the South Pole Telescope (SPT), powerful radio flares are beginning to be detected, with $L_R > 10^{15.5}$\,erg/s/Hz and reaching to $L_R \sim 10^{19}$\,erg/s/Hz \citep{Naess2021, Guns2021}. Initial analyses suggest that these flares are associated with relatively young stars \citep{Naess2021}; however, proper statistical analysis awaits significantly larger samples, such as those planned by the CCAT Observatory \citep{CCAT2021} and CMB-S4 \citep{CMB2019}.

These strong flares are important not only as a diagnostic of the magnetic and dynamo processes, but also for the energetic X-ray emission and MeV particles that interact with the surrounding disk. Combined, these processes lead to chemical reactions ionizing the disk, with diagnostics that include emission in \ion{Ne}{2} \citep[e.g.][]{glassgold2007,guedel2010} and H$^{13}$CO$^{+}$ \citep{Cleeves2017}. Finally, interactions between the MeV particles and dust grains in the disk can induce nuclear reactions, leading to short-lived radionuclides \citep{leet1998, sossi2017} and heating events \citep[see discussion by][]{shu1997}.

The most powerful YSO radio flare observed to date, $L_R = 8 \times 10^{19}$\,erg/s/Hz, was detected from the Class\,II T Tauri binary JW\,566 in Orion OMC\,2/3 in the submillimeter (submm) at 450 and 850~\micron\ with the James Clerk Maxwell Telescope (JCMT) by \citet{mairs19}. Converted to X-ray luminosity using the \citet{gudel93} relation, that event produced $\sim10^{35}$\,ergs~s$^{-1}$, though it is likely that the correlation saturates at such high values. This scaled luminosity would be an order of magnitude stronger than the X-ray super-flares observed by \citet{Getman2021}, five orders of magnitude greater than the typical T Tauri star flare, and an impressive ten orders of magnitude brighter than most solar flares. The \citet{mairs19} discovery, found via a preliminary search through the JCMT Transient Survey data set, recovered this single flare detected on 2016-11-26, with a decay of 50\% during the thirty minute integration. The submm source associated with JW\,566 was not detected at any other epochs. The spectral index between the two submm wavelengths, $\alpha = 0.11$, is broadly consistent with non-thermal emission. Together, the short submm duration and low spectral index support the flare interpretation, with the brightening event likely due to magnetic reconnection energizing charged particles to emit gyrosynchrotron/synchrotron radiation. Along with the JW\,566 burst having the largest radio luminosity, it is also unique by being observed at the highest frequency, 650~GHz (450~\micron), of any YSO radio flare to date.

In this paper, we analyze 5.5 years of JCMT Transient Survey submm monitoring observations toward eight Gould Belt star-forming regions to search for evidence of variability from faint sources, primarily Class\,II YSOs. In Section~\ref{sec:observations}, we describe the JCMT Survey, the standard data reduction procedure, and the additional processing techniques required by this paper. We present the results of our variability investigation in Section~\ref{sec:analysis} and discuss the implications of our results in Section~\ref{sec:discussion}. The paper is summarized in Section~\ref{sec:conclusions}.

\section{Observations}\label{sec:observations}
\subsection{JCMT Transient Survey}\label{sec:obs:JTS}

The JCMT Transient Survey has been monitoring eight active Gould Belt star-forming regions within $500$\,pc of the sun \citep{herczeg2017,mairs2017} since December 2015. In this paper we present an analysis of observations taken over 5.5 years, through June 2021. The survey is the first dedicated long-term monitoring program of YSOs at submm wavelengths. Each region is observed with at least a monthly cadence\footnote{Occasionally, regions are monitored with a higher cadence. For example Serpens\,Main, which hosts the 18 month quasi-periodic submm source EC 53 \citep{yoo2017,yhlee20}, also known as V371 Ser, is typically monitored with a 2-week cadence.}, when available. Almost 300 bright ($>140$\,mJy/bm) submm peaks across all regions have been investigated for variability \citep{mairs2017GBSTrans,johnstone2018,leeyh2021}, with greater than 20\% of the 83 monitored protostars showing robust evidence for long-term brightening or dimming associated with accretion processes within the disk. None of the bright protostars show evidence of single-epoch enhanced variability, and none of the submm bright starless cores are found to be variable with either short or long timescales \citep{leeyh2021}.

The eight monitored star-forming regions, namely IC\,348, NGC\,1333, OMC\,2/3, NGC\,2024, NGC\,2068, Ophiuchus, Serpens\,Main, and Serpens\,South (see Table \ref{tab:regions}) -- were selected to maximize the number of deeply embedded, Class\,0/I sources. JCMT Gould Belt Survey \citep{WardThompson2007} 850~\micron\ co-added images were used to locate bright submm peaks and collated against Spitzer YSO survey catalogues at mid- through far-IR \citep{megeath2012, dunham2015}. Along with the hundred odd protostars within these regions, more than 1500 known Class\,II YSOs (and many more Class\,III YSOs) are monitored by the survey \citep{herczeg2017}, although almost all the Class\,II systems are too faint in the submm to be detected in a single epoch. The monthly monitoring cadence was chosen based on an estimate of the thermal equilibration time for dust in a protostellar envelope coupled with the light propagation time through the envelope \citep{johnstone2013}.

\subsection{Map Reconstruction}\label{sec:obs:dr}

SCUBA-2 is a 10,000 pixel detector, simultaneously observing a 45~arcmin$^2$ footprint at 450 and 850~\micron\ \citep{holland2013}. We obtain images of 30-arcmin circular fields using the PONG1800 mode, which moves SCUBA-2 in a rotating `pong' pattern to obtain circular maps with consistent noise properties.  Each total integration is $\sim 20$ to 40 min, set based on the precipitable water vapor to reach a noise level of $\sim 12$ mJy/beam at 850~\micron.  The 8 star-forming regions in our sample have now been observed at 850~\micron\ for almost one full day (see Table~\ref{tab:regions}).

The raw instrumental observations at $850$\micron\  are affected by the ``scan-synchronous'' low frequency correlation with the telescope motion (common-mode) and by a combination of elevation induced changes in sky brightness and magnetic field pickup across the detectors \citep[see][for details]{Chapin2013}. In addition, there is attenuation due to the atmospheric opacity, flat field corrections, sky noise, cosmic rays, and other contaminants that are considered and removed by the mapmaker. The \textsc{Makemap} algorithm, explained in detail by \citet{Chapin2013} and provided as a part of the \textsc{Starlink} software \citep{Currie2014}, is employed to produce astronomical maps from the raw data. At 850~\micron, a 3\arcsec\ pixel scale is chosen in order to subsample the $\sim14\arcsec$ beam. In the pre-processing stage of the data reduction process, flat-field correction, time series down-sampling, step-correction, and de-spiking of the inputs are applied. This is followed by an iterative process, where common-mode removal, extinction correction, high-pass filtering, map estimation and white noise measurements are performed. The reduction algorithm builds the map by calculating at each pixel location the flux density (hereafter brightness) and measurement uncertainty, based on the measured signal and noise properties of the relevant bolometers. Finally, spatial filtering to suppress signal on scales $>200\arcsec$, is applied to each JCMT Transient map reconstruction to optimize the extraction of compact sources \citep[for additional details see][]{Chapin2013, mairs2017}. 

The inherent $2\arcsec$ to $6\arcsec$ pointing uncertainty of the JCMT \citep{mairs2017} is corrected by using a cross correlation technique (see Mairs et al.\ in preparation), allowing for sub-arcsecond relative alignment accuracy across epochs. Each aligned map is smoothed using a $6\arcsec$ (2 pixel) FWHM Gaussian kernel in order reduce pixel noise, slightly broadening the 850~\micron\ FWHM beam to 15\arcsec.  The relative brightness is calibrated by tracking the peak brightness of a large sample of submm sources in each region over all epochs and weighting their contributions to the calibration normalization by their individually measured signal to noise (see Mairs et al.\ in preparation). The alignment and relative calibration techniques achieve better than 2\% accuracy in the relative calibration at 850~\micron \citep[Mairs et al.\ in preparation,][]{mairs2017}, an improvement on the nominal brightness calibration of 8\% for standard reductions \citep{mairs2021}.

\subsection{Normalized Standard Deviation and Normalized Epoch Residuals}
\label{sec:obs:res}

Our goal in this paper is to search for objects that are bright in a single epoch. To discover these flares, we first measure a standard deviation map that describes the noise at every pixel location in the field. We then compare each map of the region by its standard deviation map to identify outliers as possible transient objects.

The standard deviation of the brightness at any pixel location, calculated by considering all epochs, can be compared against the expected measurement uncertainty at that location to search for unknown variable sources. In practice, in order to properly account for the significant increase in measurement noise toward the edge of each region, we use the square root of the average of the squares of the epoch-specific uncertainty measurements at the fixed location, derived during the map-making process (see Section~\ref{sec:obs:dr}). With such an approach, and assuming Gaussian statistics, the mean of the normalized standard deviation map tends to unity and the range of normalized values is inversely proportional to the root of the number of included epochs. Equivalently, the histogram of the values in the normalized standard deviation map can be approximated by a Gaussian with a peak at one and a width $N_{\rm epochs}^{-1/2} \sim 0.15$, for $N_{\rm epochs} = 30$--70. Thus, for the regions analysed here a 5-sigma outlier in the normalized standard deviation map should have a pixel value of 1.8, whereas a 25 sigma outlier will have a pixel value close to 5.

When analysing the normalized standard deviation maps, localized sources with significant enhanced variability above the measurement noise will be evident by eye or can be detected by automated routines as described in Section~\ref{sec:analysis:all}. Care must be taken, however, when searching intrinsically bright areas within each map, as the calibration uncertainty becomes comparable in magnitude to the measurement noise for sources brighter than 500\,mJy/bm. Furthermore, the JCMT focus is not always sharp and thus the 850~\micron\ beam sidelobes can produce excess emission up to distances $\sim 30$\arcsec\ from bright peaks \citep[for details on the properties of the JCMT beam see][]{mairs2021}. Given that the individual epochs are brightness calibrated using the peak brightness of known sources, this excess focus uncertainty mostly adds positive signal to the map, leading to a noise component slightly asymmetric around zero. To account for brightness calibration and focus issues, we divide each star-forming region into two areas depending on whether the mean brightness, averaged over all epochs, lies above or below $100\,$mJy/bm (equivalently, 8 times the single epoch uncertainty of $12\,$mJy/bm). Typically, only a few percent of the area within each region lies above this threshold (see Table \ref{tab:stats}).

\begin{figure*}[tbh]
\centering
\includegraphics[width=0.99\textwidth]{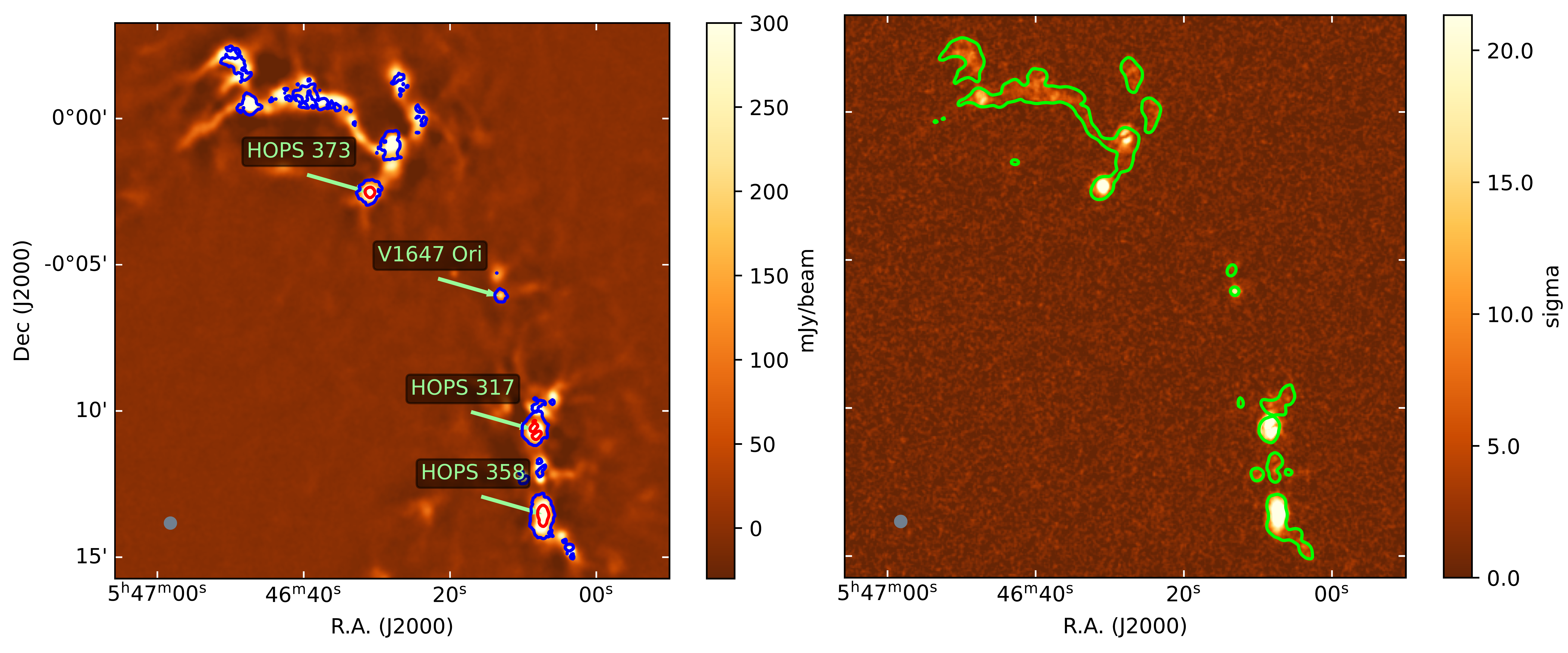}
\caption{A subsection of the NGC\,2068 region in Orion. Left: 850~\micron\ mean brightness map, combining all 52 monitored epochs. Right: normalized standard deviation map. In the right panel, the green contour traces the location where the mean brightness reaches 100\,mJy/bm. Note that within these zones the normalized standard deviation is often enhanced over the rest of the map. In the left panel, the blue (red) contour traces where the normalized standard deviation is greater than five (twenty-five) sigma above the mean. Beam-sized circular peaks in the normalized standard deviation map are evident at the locations of three known bright protostars, V1647\,Ori, HOPS\,358 and HOPS\,373 and an arc-like feature, associated with beam shape variations is seen for HOPS\,317. Additional complex features are found in the normalized standard deviation map within 30\arcsec\ of all bright structures. The 100\,mJy/bm contour effectively confines this enhanced measurement uncertainty.}
\label{fig:ngc2068}
\end{figure*}

Figure \ref{fig:ngc2068} presents the mean brightness and normalized standard deviation maps for a subsection of the NGC\,2068 star-forming region in Orion. The green 100\,mJy/bm contour marking the boundary between high and low mean brightness areas is over-plotted on the normalized standard deviation map (right panel), showing that the anomalously high standard deviation measurements are almost entirely confined to these bright regions within the map. Similarly, estimated five sigma anomalous normalized standard deviation contours are over-plotted on the mean brightness map (left panel) in blue. Red contours denote regions of extreme, 25 sigma, normalized standard deviation, highlighting the protostars HOPS\,373 to the north and HOPS\,358 to the south, as well as beam focus complications towards a bright non-variable protostar, HOPS\,317. These bright sources are best analysed directly via their light curves. The multi-wavelength time variability of HOPS\,373 has been investigated in detail by \citet{yoon2022}, while \citet{leeyh2021} found both HOPS\,373 and HOPS\,358 to be robust submm variables.

When searching for sources that vary in only one epoch, such as rare flaring events, the normalized standard deviation map is not exactly the most appropriate tool, because individual events themselves have a disproportionate effect on the standard deviation. For such events it is more effective to subtract each individual epoch from the mean brightness determined using all {\it other} epochs, i.e.\ excluding the epoch of interest, to produce a single epoch residual brightness map. To quantify the significance of residual map peaks, we again apply a normalization by dividing the residual at each pixel by an expected measurement uncertainty. In this case there are two relevant contributions to the measurement uncertainty: (1) epoch-specific uncertainties that depend on the particulars of the observing conditions and (2) location-specific uncertainties which account for both intrinsic, on-going, source variability and anomalous measurement uncertainties due to, for example, nearby sources. We again take the epoch-specific uncertainty measure directly from the map-making process (see Section \ref{sec:obs:dr}) and here we equate the location-specific measure with the standard deviation of the brightness at each pixel across all other epochs. We then normalize each single epoch residual map by the {\it larger} of these two uncertainty values. As a guide, in Table \ref{tab:regions}, the epoch-specific measurement uncertainty ranges for each region are presented.  For this calculation, only those areas lying below the $100$\,mJy/bm contour and within the central 15\arcmin\ radius of each region are utilized, ensuring that the time integration per pixel is uniform.
 
These analyses were performed on the full 20--40 minute integrations. Often flare lifetimes may be shorter. The $5.5\sigma$ detection limit of 65\,mJy/beam corresponds to the summed integration, so a flare that lasted only $\sim10$~min would need to have a peak of $>180$\,mJy/beam to be identified. Flares that last for only a minute may be missed entirely, since each region in a map is only visited $\sim$ 10 times during a single observation \citep{mairs19}.

\begin{figure*}[tbh]
\centering
\includegraphics[width=0.99\textwidth]{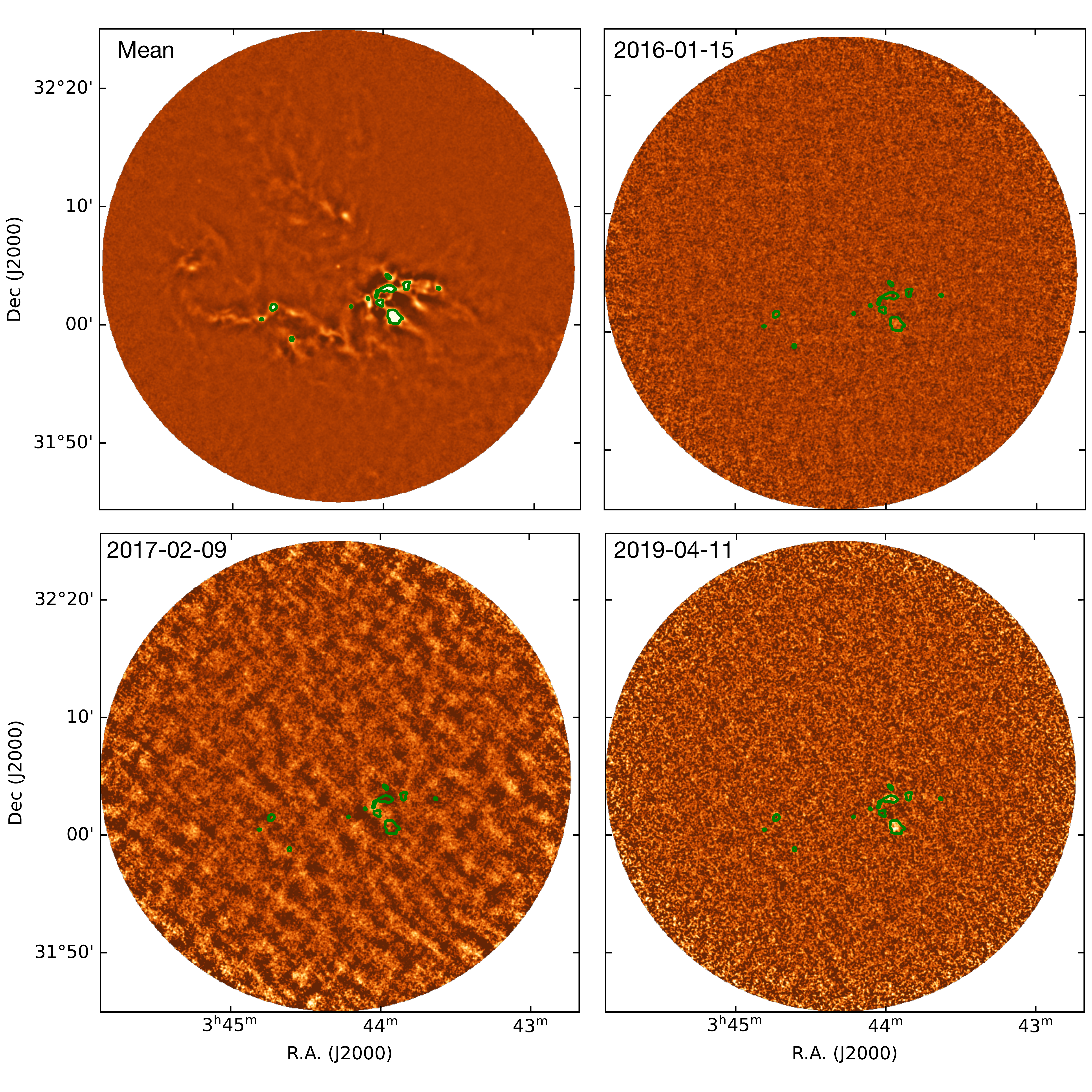}
\caption{The IC\,348 region in Perseus. Top Left: 850~\micron\ mean brightness map, combining all 34 monitored epochs ($\sigma = 2$\,mJy/bm). Top Right: residual map for epoch 2016-01-15 ($\sigma = 11$\,mJy/bm). Bottom Left: residual map for epoch 2017-02-09 ($\sigma = 11\,$mJy/bm). Bottom Right: residual map for epoch 2019-04-11 ($\sigma = 16\,$mJy/bm). All residual maps are plotted over the same range of brightness values. In each panel the green contour traces the location where the mean brightness reaches $100\,$mJy/bm. Most epochs have spatially uniform noise uncertainties similar to epoch 2016-01-15, although in poor weather the normalization may be worse as demonstrated by epoch 2019-04-11. Rarely, more complex spatially correlated noise is found in the residual map, as shown for epoch 2017-02-09.}
\label{fig:ic348}
\end{figure*}

Figure \ref{fig:ngc2068} clearly shows the importance of location-specific measurement uncertainties within each star-forming region, especially in areas of high brightness. Figure \ref{fig:ic348} presents example single epoch residual maps of IC\,348, revealing the variety of possible epoch-specific measurement uncertainty categories. While the majority of epochs are similar to 2016-01-15, where the measurement noise has $\sigma = 11\,$mJy/bm and is uniformly distributed, very occasionally an epoch will have anomalous spatially-correlated noise similar to 2017-02-09 but where the measurement noise remains low, $\sigma = 11\,$mJy/bm. Somewhat more common are epochs, such as 2019-04-11, where the noise properties remain uniformly distributed but are enhanced due to instrument or weather conditions; in this extreme case $\sigma = 16\,$mJy/bm. Given this knowledge, after determination of candidate single epoch variables, it is important to check for potential measurement noise anomalies in the observed epoch.

\section{Analysis}\label{sec:analysis}
\subsection{Searching for Faint Variables Over All Epochs}
\label{sec:analysis:all}

\begin{figure*}[tbh]
\centering
\includegraphics[width=0.99\textwidth]{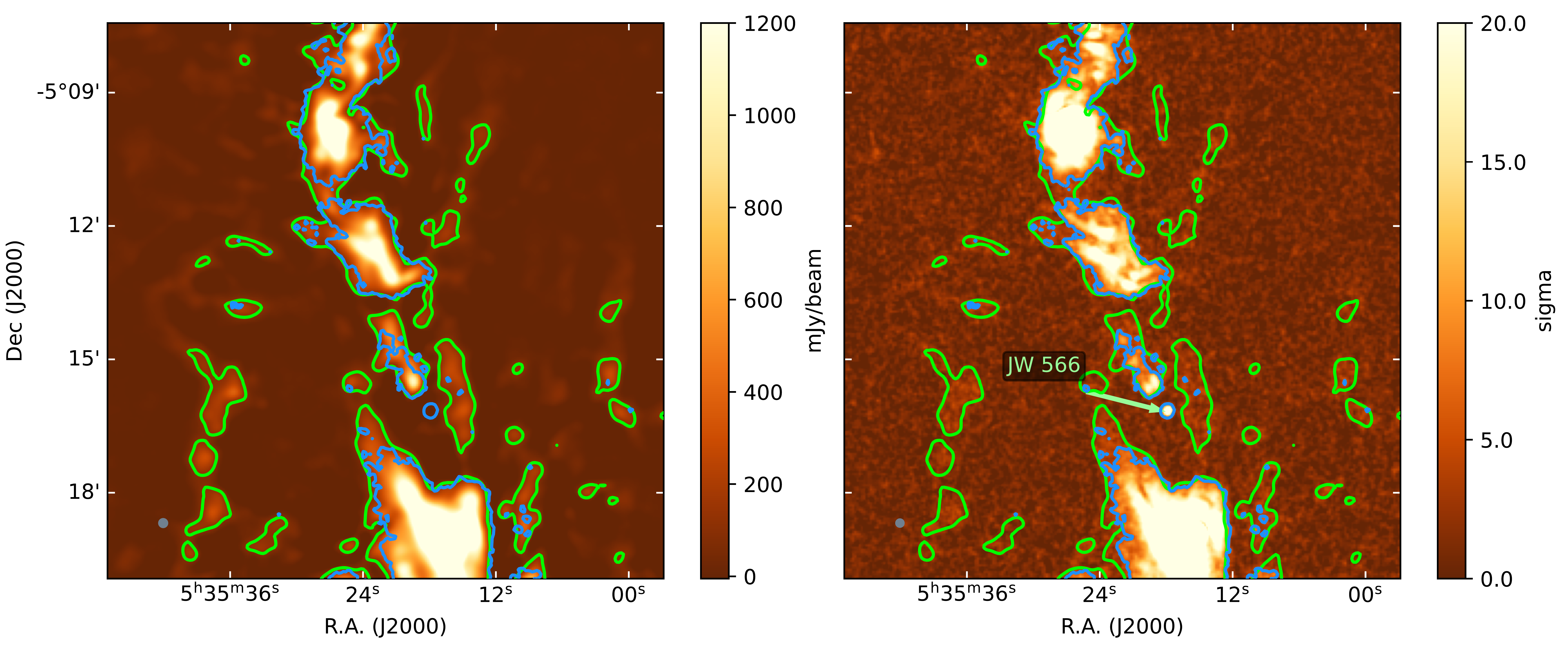}
\caption{A subsection of OMC\,2/3 in Orion. Left: 850~\micron\ mean brightness map, combining all 42 monitored epochs ($\sigma = 2$\,mJy/bm). Right: normalized standard deviation map over all monitored epochs.
In both panels the green contour traces the location where the mean brightness reaches 100\,mJy/bm and the blue contour marks where the normalized standard deviation reaches five sigma above the mean. JW\,566 stands out clearly in the normalized standard deviation map but is too faint to be observed in the mean brightness map despite the exquisite map sensitivity.}
\label{fig:omc23}
\end{figure*}

\begin{figure*}[tbh]
\centering
\includegraphics[width=0.99\textwidth]{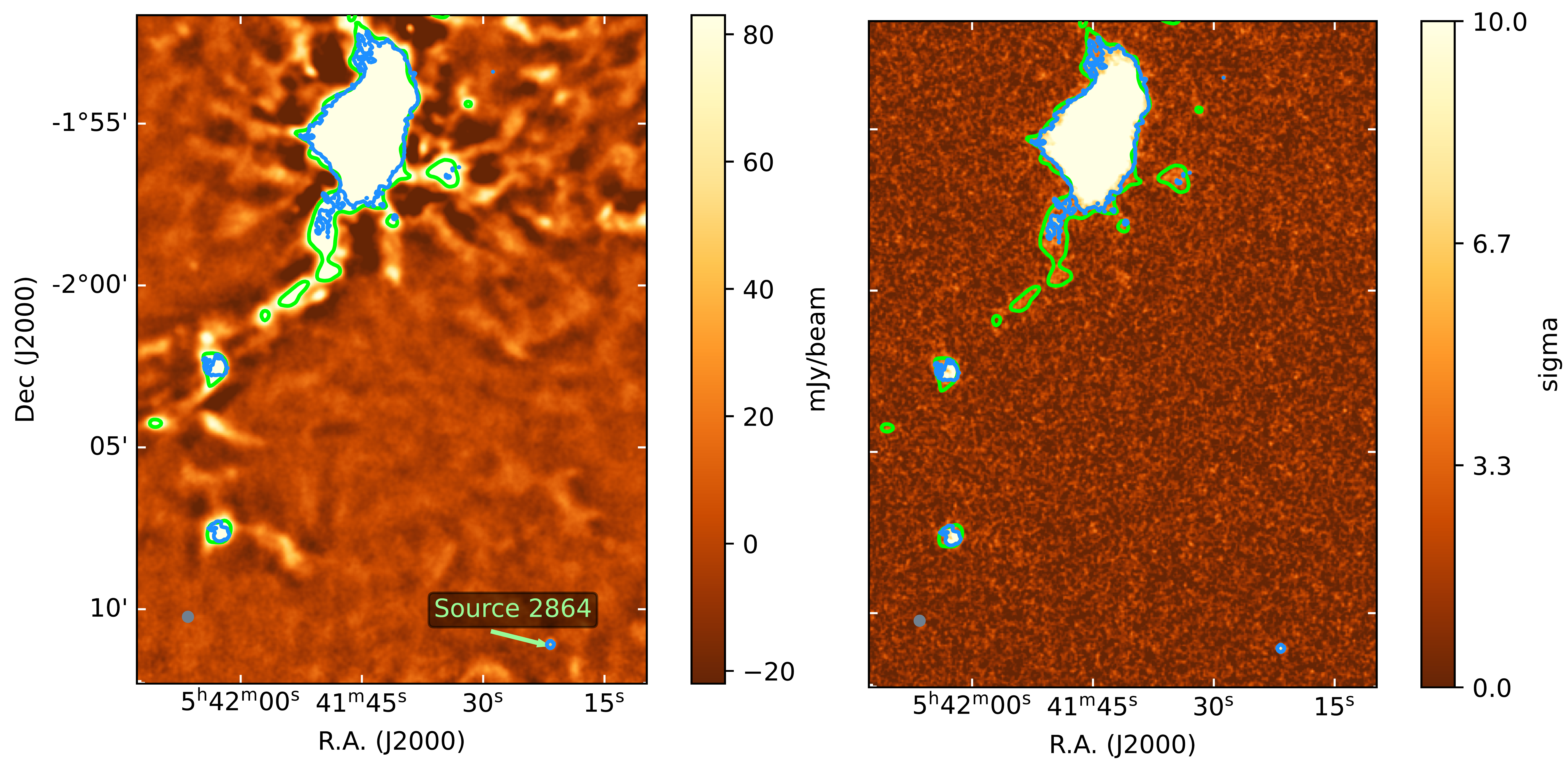}
\caption{A portion of the NGC\,2024 region in Orion, with NGC\,2023 just off the bottom of the mapped area. Left: 850~\micron\ mean brightness map, combining all 43 monitored epochs ($\sigma = 2$\,mJy/bm). Right: normalized standard deviation map over all monitored epochs. In both panels the green contour traces the location where the mean brightness reaches $100$\,mJy/bm and the blue contour marks where the normalized standard deviation reaches five sigma above the mean. A single object, Source 2864, stands out at the bottom right of the normalized standard deviation map. While faint, this object is also observed in the mean brightness map.}
\label{fig:n2024}
\end{figure*}

As a first step to identifying variable sources, the normalized standard deviation maps for each region produced using all available epochs were analysed by eye. For this analysis the 100\,mJy/bm contour was used to mark the bright areas, with their known higher residual uncertainty, as distinct from the bulk of each map where the noise properties are significantly more uniform. Although not the focus of this paper, within the bright areas of these star-forming regions we often find circularly shaped localized residuals at the locations of embedded protostars. In all cases, these sources were previously known to vary in brightness and are more effectively investigated via their light curves \citep{johnstone2018,leeyh2021}. The confined bright areas also present arc-like residuals tracing the extended sidelobes of the telescope beam due to focus issues (see for example the location of HOPS 317 in the right panel of Figure~\ref{fig:ngc2068}). Alternatively, within the low brightness areas of these star-forming regions, by eye we find only two locations with significant enhanced normalized standard deviation measurements. These locations are coincident with JW\,566 in OMC\,2/3 (Figure~\ref{fig:omc23}), previously discovered as a transient source by \citet{mairs19}, and a Spitzer-identified candidate Class\,II YSO in NGC\,2023, source MGM12\,2864 (hereafter Source\,2864) in the catalogue by \citet[MGM12]{megeath2012}, at the edge of the NGC\,2024 map (Figure~\ref{fig:n2024}).

An automated procedure was used to search each star-forming region for evidence of anomalous standard deviation measurements at the locations of the known YSOs. For this analysis we utilized the Spitzer-based YSO catalogues by \citet{megeath2012}, \citet{stutz2013}, and \citet{dunham2015}. In Table \ref{tab:stats} we count the YSOs within the faint brightness area of each region, where the number ranges from 56 (Serpens\,Main) to 358 (OMC\,2/3). These YSOs occupy a small, typically few percent, fraction of each region's area (see Table \ref{tab:stats}). Thus, the range of normalized standard deviation measures found within each map can be used to determine the likelihood that a large value found at the location of a YSO is due to intrinsic source variability. 

In detail, a roughly beam-sized box of 5 by 5 pixels {(15\arcsec\ by 15\arcsec)} was centered on each YSO location {to ensure that the peak was well sampled despite a several arcsecond uncertainty in absolute image registration}. The fifth largest pixel value in the box was used as a proxy for the source variability in order to minimize the shot noise associated with individual pixels {due to the pixel gridding oversampling the beam}. This value was then compared against both the total number of pixels higher than that value across the rest of the normalized standard deviation map, excluding areas with bright emission, and the fractional area of the map containing YSOs to estimate the likelihood of such an enhanced YSO brightness deviation measurement within the region being due to random chance. The False Alarm Probability (FAP), therefore, decreases when the fifth largest pixel value in a given YSO increases and as the number of known YSOs within a region decreases. 

We set a relatively shallow FAP threshold of 10\% for candidate faint variables within each individual region, {in order to test for variables near the sensitivity limit}, yielding an expectation of one potential false alarm over the eight analysed regions. In total, seven candidates were recovered by this process. Five of the candidates were  subsequently withdrawn after being found to lie extremely close to the $100\,$mJy/bm contour where the residual uncertainty could be seen to  spatially spread. The only two robust detections via the automated process are the same sources as found by eye, JW\,566 in OMC\,2/3 (FAP$< 0.01$\%) and Source\,2864 in NGC\,2023 (FAP$<3$\%).
     
\subsection{A Blind Search for Transient Variables by Epoch}
\label{analysis:epoch}

Following the approach developed in Section~\ref{sec:analysis:all}, for each region the individual epochs were analysed carefully to search for transient events localized in time. Here, rather than using the normalized standard deviation map to uncover potential variables, we instead analysed the normalized residual signal in each epoch, subtracting off the mean of all other epochs and dividing by the expected measurement uncertainty at each pixel (see Section~\ref{sec:obs:res}). For this analysis we exclusively consider the areas within each region that lie below the 100\,mJy/bm contour in order to avoid complications due to the brightness calibration and telescope focus.

Each star-forming region map consists of half a million pixels, or about twenty thousand independent beams.\footnote{Recall that the map pixels are 3\arcsec\ in length and the smoothed 850~\micron\ JCMT beam has a FWHM of $\sim15\arcsec$.} Thus, a greater than four sigma likelihood result is expected about once per epoch due to random chance alone. Given that 34-67 epochs are observed for each of the eight regions, at least one five sigma false positive is expected to be found in the full sample, assuming Gaussian statistics. Thus, we first searched the normalized residual maps for each epoch of each region for anomalous residual peaks higher than 5.5 sigma. Such events were found in four epochs, specifically IC\,348 (2015-12-22, 2017-02-09), NGC\,1333 (2015-12-22), and OMC\,2/3 (2016-11-26). Inspection of the specific images, however, revealed that three of these maps contained correlated noise resulting in extended blooms across the map and leaving enhanced normalized residuals (see for example the bottom right panel in Figure~\ref{fig:ic348}, IC\,348 epoch 2017-02-09). Only epoch 2016-11-26 of OMC\,2/3 (see Figure~\ref{fig:omc23e}) contains a viable peak greater than 5.5 sigma, at the location of JW\,566 T Tauri star \citep[as already reported by][]{mairs19}. Other than JW\,566, to date within the JCMT Transient Survey there are no epoch-specific brightening events above an upper limit of 5.5 sigma. Given that the typical epoch-uncertainty is 12~mJy/bm (Table~\ref{tab:regions}) this is equivalent to $\sim 65$\,mJy/bm at 850~\micron.  

\subsection{Searching for Single-Epoch Transients of Known YSOs}

\begin{figure*}[bt]
\centering
\includegraphics[width=0.99\textwidth]{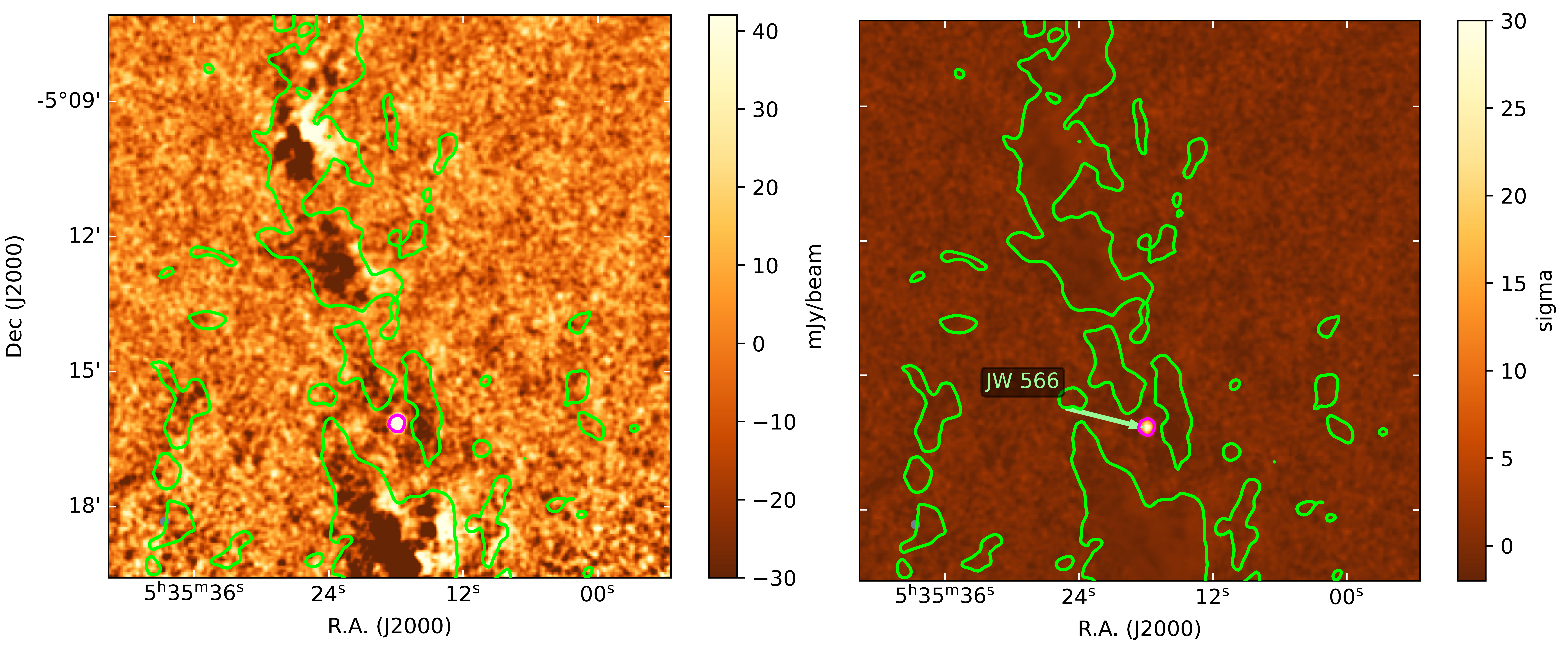}
\caption{A subsection of OMC\,2/3 in Orion. Left: 850~\micron\ residual brightness map after subtracting the mean brightness over all other epochs from epoch 2016-11-26. Right: normalized residual map for epoch 2016-11-26. In both panels the green contour traces the location where the mean brightness reaches 100\,mJy/bm and the magenta contour marks where the normalized residual map reaches five sigma above the mean. JW\,566 stands out clearly in both the residual map and the normalized residual map. These images should be compared against Figure \ref{fig:omc23}.}
\label{fig:omc23e}
\end{figure*}

At the locations of the known infrared YSOs in each region (Table \ref{tab:stats}) it is possible to search deeper within the normalized residual maps, using the procedure outlined in the previous section whereby we compare the single epoch normalized residual measurements obtained at the location of the known sources against the distribution of measurements across the larger map. Again, we search for the fifth largest pixel value within a 5 by 5 pixel region centered on each known YSO and compare against both the number of pixels with this value or higher in the full map and the fraction of the map covered by YSOs. We consider candidate YSO transient events detected in a given epoch when the FAP per epoch is less than 10\%. Even with this low detection threshold, we find only seven candidate events within the almost 400 JCMT Transient Survey epochs. This is somewhat less than the 40 events expected solely due to the FAP threshold, suggesting that the locations of YSOs are slightly less variable than the rest of each map (potentially due to leakage of excess uncertainty across the 100\,mJy/bm boundary). Furthermore, all of the seven candidate events have peak residual normalized signal below 4.5 sigma (see Table~\ref{tab:stats}) with the exception of JW\,566, which reaches 25.9 sigma (Figure~\ref{fig:omc23e}). The large increase in the significance of the JW\,566 detection due to the epoch-specific analysis versus the all-epoch analysis is expected because the extraordinary brightening event is {\it removed} from the epoch-specific determination of the expected standard deviation across epochs. Similarly, the lack of detection of significant variability from Source\,2864 in NGC\,2023 via the epoch-specific analysis suggests that it did not undergo a single transient event but rather it is a faint long-term submm variable source. We discuss further the enigmatic Source\,2864 in Section \ref{sec:disc:blazar}.

Across eight star-forming regions, the JCMT Transient Survey covers 1200 YSOs (Table \ref{tab:stats}). Furthermore, each region has about 40 epochs to date (Table \ref{tab:regions}). In total, roughly 50,000 individual YSO 850~\micron\ brightness measurements and single epoch normalized residuals have been made by the JCMT Transient Survey. Assuming Gaussian statistics, we therefore expect at least one $\sim4.5$ sigma event just by random chance. As such, the YSO candidates identified above with peaks lower than 4.5 sigma are not robust. With the exception of the extraordinary bursting event coincident with JW\,566, we therefore assert that there are no epoch-specific YSO brightening events above an upper limit of 4.5 sigma, $\sim 55$\,mJy/bm at 850~\micron\ in the JCMT Transient Survey data set.

\section{Discussion}\label{sec:discussion}
\subsection{Submm Flaring Events from YSOs}
\label{disc:flares}

\citet{mairs19} discovered an extreme flaring event from the T Tauri star JW\,566 within the JCMT Transient Survey monitoring of OMC\,2/3.  We also recover JW\,566 through our more detailed analysis over all regions and epochs to date. Our search of known, infrared-selected YSOs reveals no additional detections of single-epoch brightening events larger than 4.5 sigma, i.e., $\sim 55$\,mJy/bm, or equivalently a factor of 8 fainter in brightness than JW\,566. Our blind analysis also reveals zero other single-epoch transient events larger than 5.5 sigma, or 65~mJy/bm.

This robust non-detection is surprising given the expectation that flaring events, like most stochastic distributions, are thought to follow a power-law distribution with a greater number of events at lower brightness. For example, \citet{Getman2021} find that dN/d$L_X \propto {L_X}^{-\alpha}$, with $\alpha \sim 2.1$ for both diskless and disk-bearing pre-main sequence stars, when $L_X > 10^{32.5}$erg/s. Furthermore, the JW\,566 event was observed to decay by $\sim50$\% over the half hour epoch \citep{mairs19}, which implies that it would have remained observable above our $\sim 55\,$mJy/bm threshold for at least an hour, assuming a linear decay, or two hours, if the decay were exponential. Taken together, these arguments imply that one should expect more fainter detections than bright ones. We therefore conclude that the JW\,566 submm flaring was an exceptional rare or unique event, and that its occurrence rate within the JCMT Transient Survey is not directly related to an underlying power-law distribution of bursts.

The JCMT Transient Survey has monitored about 1200 infrared-selected YSOs for about 40 epochs each, equivalently 50000 individual half-hour observations. Additional Class\,III (diskless) YSOs are likely present in the field, a missing and older population that is often revealed through analysis of Gaia astrometry.
This represents about 1000 days, equivalently 3 years, of 850~\micron\ monitoring of a random YSO. Taking our non-detection threshold at $\sim 55$\,mJy/bm and assuming a distance of 400\,pc for the typical monitored YSO, this converts to a radio luminosity threshold of $L_R \sim 10^{19}$\,erg/s/Hz, above which we assert that there have been no YSO radio flaring events detected, other than the exceptional source JW\,566. Assuming the \citet{gudel93} scaling relation between radio and X-ray luminosity, yields a lower limit x-ray luminosity of $L_X \sim 10^{34}$\,erg/s. This remains an extremely high threshold, on par with the brightest x-ray flares in the super-flare sample by \citet{Getman2021} and likely at the limit where X-rays are saturated and cannot further increase in luminosity. The nearest star-forming region within the JCMT Transient Survey, Ophiuchus, is approximately three times closer than this nominal distance. For this region, the 100 YSOs, monitored over 34 epochs and corresponding to 70 days of observation of a single YSO, yield an upper limit on radio luminosity from flares of $L_R \sim 10^{18}$\,erg/s/Hz (or $L_X \sim 10^{33}$\,erg/s).

{As discussed in Section 2.1, radio flares may vary on timescales shorter than the standard half-hour JCMT epoch integrations. Thus,  care must be taken in considering the limits presented here - which assume a constant brightness throughout the observation - when comparing with models or other radio surveys. Furthermore, while each epoch integration is approximately thirty minutes, the detector array is substantially smaller than the image field of view and thus the brightness measurement at any given location is an average over a summation of shorter, time-separated integrations \citep[for details see][]{mairs19}.}

As noted in the introduction, all-sky mm surveys with ACT and SPT \citep{Naess2021,Guns2021} have caught a handful of radio flares from young stars at luminosities close to the non-detection threshold presented here. Planned campaigns by the CCAT-prime collaboration \citep{CCAT2021} and the CMB-S4 consortium \citep{CMB2019} should add significantly to this sample. JW\,566, however, remains almost an order of magnitude brighter than any of these other detections. For older M-stars, mm monitoring for radio flares in the mm by \citet{Macgregor2018} and \citet{Macgregor2020} has uncovered small flares but no superflares. Finally, we note that the JW\,566 flare is doubly remarkable due to both its extreme radio luminosity and its remarkably high radio frequency, the 650\,GHz measurement being almost an order of magnitude higher in frequency than either the ACT or SPT detections \citep{mairs19}. 

We therefore speculate that the rareness of the event may be due to a requirement that to be observed at high (submm) frequencies the event must also be extremely strong. Turning the hypothesis around, such a scenario would introduce a minimum luminosity threshold for submm flares which could explain the lack of lower brightness detections by the JCMT Transient Survey.  Observational confirmation of this scenario would require detection of additional bright flares at 650\,GHz, similar to that of JW\,566, while no fainter events are detected. 

Interestingly, JW\,566 is a known binary system \citep{daemgen2012} with a disk around at least one component, as seen at mid-IR and mm wavelengths \citep{megeath2012,hacar2018}. This might increase the complexity of the magnetic structure of the system and its interaction with the surrounding disk material. Indeed, radio synchrotron emission has been mapped in the hierarchical system V773\,Tau~A \citep{massi2006, torres2012}, which has four components, including a tight inner binary. The enhanced activity has been shown to be produced by interacting helmet streamers and at periastron the radio brightness can be raised to more than 30 times that at apoastron \citep{adams2011}. The spectroscopic binary DQ\,Tau also exhibits strong mm flares that erupt at periastron \citep{salter2010}.
Unlike V773\,Tau~A or DQ\,Tau, JW\,566 has only produced one observable powerful flare thus far, however, it remains intriguing to speculate that this single event highlights JW\,566 as having an unusual system architecture, perhaps with at least one component also being a tight and yet-unresolved binary.  The rarity of this event may be partially explained if such bright flares occur only during periastron passage of close binaries, which are $\sim 3$\% of the total population \citet{mazzola20,kounkel21}.

\subsection{An Enigmatic Blazar Masquerading as a YSO Candidate}
\label{sec:disc:blazar}

Along with JW\,566 in OMC2/3, our search for faint transient events uncovers a multi-epoch variable near the southern edge of the NGC\,2024 star-forming region field, at R.A.\ 05:41:21.7, Dec.\ $-02:11:08.3$, associated with Source\,2864 in \citet{megeath2012} and also GBS-VLA J054121.69$-$021108.3. This source is likely a blazar, as indicated by a parallax consistent with a background object \citep{kounkel17}, but it has at times been previously classified as a Class\,II YSO \citep{megeath2012} due to its spectral energy distribution and its location within an active star-forming region and rejected as a quasar because of the YSO classification.

\begin{figure*}[bt]
\centering
{
\includegraphics[width=0.99\textwidth]{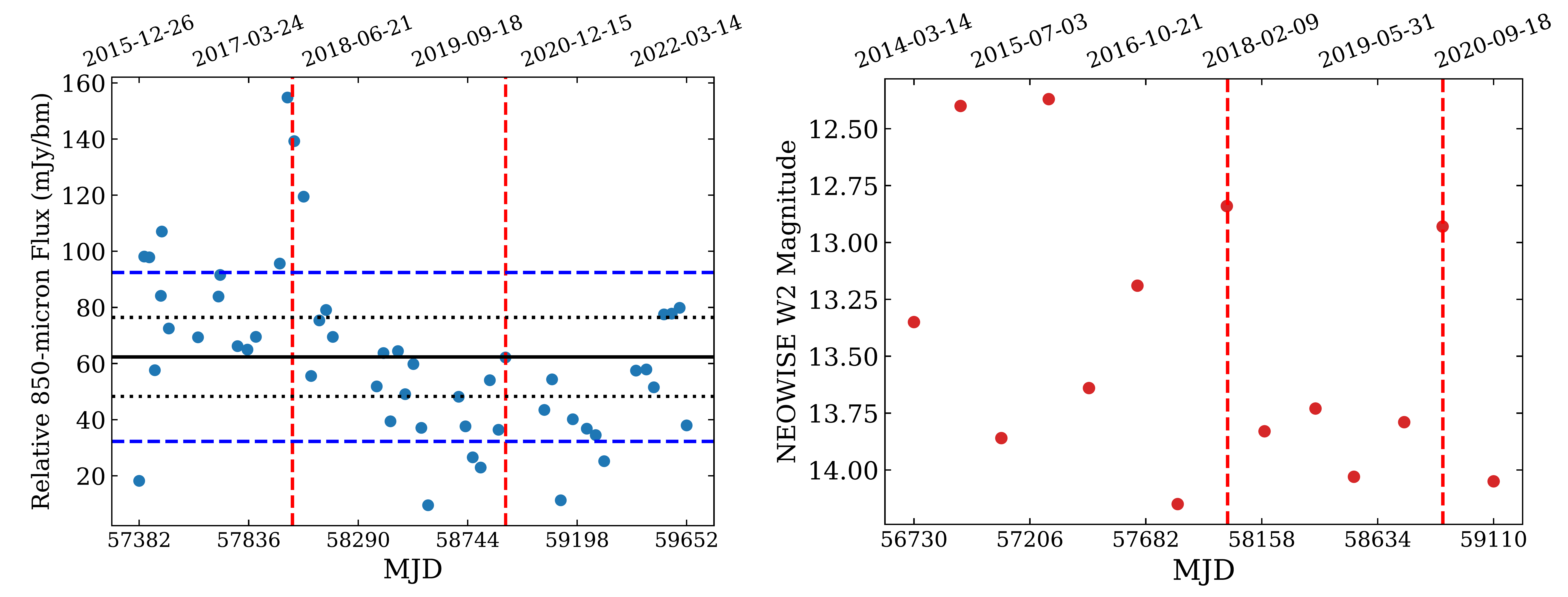}
}
\caption{
Left: JCMT submm 850~\micron\ lightcurve of Source\,2864 in NGC\,2023. The solid horizontal line shows the mean flux over all epochs while the dotted horizontal lines indicate the expected uncertainty due to measurement noise. The blue dashed horizontal lines indicate the measured uncertainty around the mean. Right: NEOWISE $W2$ 4.6\micron\ lightcurve of Source\,2864 in NGC\,2023. In both panels the red dashed vertical lines denote two mid-IR brightening events observed by NEOWISE (see text).}
\label{fig:107_2864}
\end{figure*}

In contrast with JW\,566, Source\,2864 is detected in the every epoch of submm monitoring of the field. The left panel of Figure~\ref{fig:107_2864} reveals a dimming over time, with a short months-long burst in Autumn 2017. Over all epochs, the mean peak brightness at 850~\micron\ is $S = 53$\,mJy/bm while the variation around the mean is $\sigma = 28$\,mJy/bm, a factor of two larger than the calculated measurement uncertainty in the immediate vicinity of the source, 12--15\,mJy/bm.\footnote{As Source\,2864 is near the edge of the NGC\,2024 field, the surrounding measurement uncertainties are somewhat larger than the typical values of 11--12\,mJy/bm presented in Table~\ref{tab:regions}.} This source is fainter than the brightness limits for earlier JCMT-Transient analyses \citep{mairs2017GBSTrans,johnstone2018,leeyh2021}. Nevertheless, Source\,2864 has a fractional variability, $\sigma/S \sim 0.5$, which is similar to the strongest submm variables detected in those analyses. 

Source 2864 has been previously observed and classified using near through mid-IR measurements. The source is not visible in the 2MASS survey at any of J, H, or K.  Based on Spitzer mid-IR photometry, \citet{mookerjea09} classified it as a highly extincted Class\,II source. This Spitzer classification was also assessed by \citet{megeath2012} based on the, non-extinction corrected, IRAC spectral index $\alpha = 0.03$. However, \citet{povich13} reconsidered the classification and calculated that almost 40 magnitudes of extinction are required to explain the optical through mid-IR spectral energy distribution, if the source is to be considered a late-stage YSO. 

The extreme extinction raises suspicions that Source\,2864 is only masquerading as a YSO. {It is well known \citep[e.g.][]{Harvey07, Gutermuth09, kryukova14} that YSOs have infrared colors similar to background galaxies and AGN \citep[for a review see][]{megeath22}. More recently, attempts to catalogue galaxies using their mid-IR colors  \citep{rakshit19, paggi20} have had to deal with YSO contamination: especially for blazars and BL\,Lac objects.} Due to its mid-IR colors, Source\,2864 was included in the initial ALMA blazar catalogue by \citet{paggi20} but subsequently culled due to its pre-existing classification as a YSO.

Indeed, many of the measured properties of Source 2864 are similar properties of young stars in Orion. For example, Source\,2864 is X-ray bright. Using ASCA, \citet{yamauchi00} found it to be the hardest X-ray source toward NGC\,2023, with an attenuation that is converted to A$_{\rm v} \sim 30$, similar to the extinction estimate from the infrared. The authors note that the hardness of the X-rays might also be explained with an extragalactic classification; however, the non-extinction corrected X-ray brightness is only slightly lower than for the other dozen X-ray sources revealed in the field. At the distance of Orion, the observed brightness is equivalent to $L_X \sim 10^{30}$\,erg/s. \citet{lopez13}, using XMM-Newton, also found Source\,2864 to be X-ray bright, with an uncorrected x-ray brightness similar to their other NGC\,2023 targets.

While the infrared and X-ray properties are consistent with either a YSO or a background blazar, the radio properties demonstrate that the blazar interpretation is correct.  Source 2864 was the brightest object in the field in a 3.6 cm survey by \citet{reipurth2004}. \citet{kounkel2014} later noted that Source\,2864 has the highest centimeter flux density reported for any young star. The authors also found it to be highly variable at radio wavelengths, with variations at almost the 50\% level. The source remained extremely bright in follow-up VLBA observations \citep{kounkel17}, requiring a very small angular size for the emission region. Additionally, the VLBI image of Source\,2864 reveals a core-jet structure \citep{petrov2021} that is typical of extragalactic sources. More importantly, the source revealed no measurable parallax, implying a location more distant than Orion. The source is therefore almost certainly a background, extragalactic, source, which, following the mid-IR color selection by \citet{paggi20}, we identify as a blazar. 

Source\,2864 is highly variable across many wavelengths - from the radio through the mid-IR (see Fig.~\ref{fig:107_2864}). Using the variability classification scheme of mid-IR NEOWISE lightcurves developed by \citet{park2021}, Source\,2864 is classified a a mid-IR irregular variable with a large observed flux standard deviation. Furthermore, across the last 5.5 yrs, Source\,2864 is dimming in the mid-IR in a similar manner to the submm.
The source also undergoes occasional brightness jumps of $\sim 1-1.5$ mag, including a mid-IR burst near MJD\,58000 almost coincident in time with the JCMT-observed submm burst. Indeed, the three JCMT data points marking the burst span from 18 days before the NEOWISE observation to 49 days after. Unfortunately, an earlier NEOWISE burst around MJD\,57250, preceded the start of the JCMT Transient Survey by 100 days and a later burst near MJC\,58900 occurred just as Orion became a daytime target at the JCMT in Winter 2020, unobservable for the next few months.

Similarity between observed mid-IR and submm lightcurves has been previously noted for protostars by \citet{contreras20}, where the expectation is that the mid-IR traces the variable accretion rate near the protostar and the submm traces the temperature response in the envelope to the accretion \citep[see also theoretical arguments by][]{johnstone2013,macfarlane2019a,macfarlane2019b,baek20}. As anticipated theoretically, \citet{contreras20} found that the typical fractional time-dependent response was much smaller in the submm than in the mid-IR, such that $F_{850} \propto F_{\rm mid-IR}^{1/5.5}$.  Source\,2864, however, appears to deviate significantly from this relationship, showing a factor of two brightness variation for the MJD\,58000 burst in both the mid-IR and submm. Furthermore, radio variability for this source has also been observed at the 50\% level \citep{kounkel2014}. We therefore speculate that multi-wavelength variability can be used to classify sources, breaking the color-color diagram degeneracy between YSOs and extreme extragalactic sources.\footnote{Periodic mid-IR lightcurves have similarly been used to separate background AGB stars from YSOs in the Gould Belt \citep{leeje2021,park2021}.} Around YSOs, the physical environments responsible for various aspects of the spectral energy distribution, from optical through radio, are distinct and therefore can be expected to respond to system changes with differing amplitudes, and possibly time delays. For extreme extragalactic sources, such as blazars, the physical process responsible for the spectral energy distribution is less clearly understood but expected to be dominated by non-thermal processes producing a more consistent variability across wavelengths, as observed here for Source\,2864. For blazars, the spectral energy distribution in the submm through mid-IR is often dominated by synchrotron emission that leads to correlated variability in this region of the spectrum \citep[e.g.][]{hartman96}.

Finally, we have analysed the JCMT submm lightcurve of Source\,2864 (Figure \ref{fig:107_2864}: Left) to search for potential secularity and inherent timescales.\footnote{Additional time-dependent mm through submm observations of Source\,2864 are available from ALMA, where it is used as an occasional secondary calibrator \citep{bonato19}. Although these data cover a range of wavelengths, they are significantly sparser than either the JCMT or NEOWISE observations and therefore not used here for time-dependent analyses.} Considering the observations prior to 2021, a robust linear fit is found with a slope of $-12\,$mJy/bm/yr, however after that date it has flattened and possibly begun to rise again. Anticipating that the light curve remains tied to an underlying mean brightness, while undergoing significant deviations, a structure function analysis \citep[see][for methods]{sergison20} was performed which found increased power on longer timescales but no clear associated intrinsic time constant. Similarly, a damped random walk analysis \citep[see][for methods]{kelly09,dexter14} was performed in order to estimate a saturation time scale (the timescale beyond which the amplitude of variability does not increase), with the results suggesting a value larger than the presently monitored 5.5 years. As shown by \citet{bower15}, such a long submm-measured saturation time is not uncommon for blazars.

\section{Conclusions}\label{sec:conclusions}

In this paper we have analyzed 5.5 years of JCMT Transient Survey submm monitoring observations toward eight Gould Belt star-forming regions to search for evidence of variability from faint sources, primarily Class\,II YSOs. Eight regions, which include infrared-selected $\sim 1200$ young stellar objects, have been monitored an average of 47 times, with each integration lasting about 30 minutes.

Here we summarize the main results of the paper:
\begin{itemize}
\item When searching the normalized standard deviation maps derived using all epochs, only two robust source detections are recovered, JW\,566 \citep{mairs19} in OMC\,2/3 (FAP$< 0.01$\%) and Source\,2864 \citep{megeath2012} in NGC\,2023 (FAP$<3$\%);

\item Other than JW\,566, to date within the JCMT Transient Survey there are no epoch-specific brightening events above an upper limit of 5.5 sigma, or $\sim 65$\,mJy/bm at 850~\micron. Furthermore, at the locations of known YSOs there are no epoch-specific brightening events above an upper limit of 4.5 sigma, $\sim 55\,$mJy/bm at 850\micron. Taking a distance of 400\,pc for the typical monitored YSO, the brightness limit above which no additional single epoch events are detected converts to a radio luminosity threshold of $L_R \sim 10^{19}$\,erg/s/Hz, and assuming the \citet{gudel93} scaling relation between radio and X-ray, yields a threshold luminosity of $L_X \sim 10^{34}$\,erg/s. This threshold lies at the top end of the super-flare X-ray luminosity sample analysed by \citet{Getman2021}.  The largest radio flares, like JW\,566, may access energies that are beyond the saturation limit of X-rays emission;

\item The JCMT Transient Survey has monitored about 1200 YSOs for about 40 epochs each, equivalent to 3 years of 850~\micron\ monitoring of a random YSO, with only a single burst event detected. The 100 YSOs in the core of Ophiuchus monitored over 34 epochs correspond to 70 days of observation of a single YSO, with no bursts detected. For the Ophiuchus sample, the upper limit on radio luminosity from flares, given our lack of detection, is $L_R \sim 10^{18}$\,erg/s/Hz and the assumed conversion to x-ray luminosity yields $L_X \sim 10^{33}$\,erg/s;

\item The powerful radio flare from JW\,566 \citep{mairs19}, a binary T Tauri star in OMC2/3, remains a unique and rare event, perhaps related to binarity;

\item  We have identified one variable quasar in approximately 1.6~sq.~deg. of monitoring.  The quasar, Source\,2864 from \citet{megeath2012}, visually coincident with NGC\,2023, is most likely a background blazar, the first extragalactic submm variable source detected by the JCMT Transient Survey. Consideration of variability strength as a function of frequency across the electromagnetic spectrum may allow for better classification of sources which overlap in the mid-IR color-color diagram, such as YSOs and extreme extra-galactic objects. 

\end{itemize}

\section*{Acknowledgements}

{The authors thank the referee for comments that improved this paper.} D.J.\ and G.J.H.\ appreciate a discussion with Marina Kounkel on the VLBI parallax non-detection of Source 2864.

The authors wish to recognise and acknowledge the very significant cultural role and reverence that the summit of Maunakea has always had within the indigenous Hawaiian community. We are most fortunate to have the opportunity to conduct observations from this mountain. 

The James Clerk Maxwell Telescope is operated by the East Asian Observatory on behalf of The National Astronomical Observatory of Japan; Academia Sinica Institute of Astronomy and Astrophysics; the Korea Astronomy and Space Science Institute; the Operation, Maintenance and Upgrading Fund for Astronomical Telescopes and Facility Instruments, budgeted from the Ministry of Finance (MOF) of China and administrated by the Chinese Academy of Sciences (CAS), as well as the National Key R\&D Program of China (No. 2017YFA0402700). Additional funding support is provided by the Science and Technology Facilities Council of the United Kingdom and participating universities in the United Kingdom and Canada. Additional funds for the construction of SCUBA-2 were provided by the Canada Foundation for Innovation. The James Clerk Maxwell Telescope has historically been operated by the Joint Astronomy Centre on behalf of the Science and Technology Facilities Council of the United Kingdom, the National Research Council (NRC) of Canada and the Netherlands Organisation for Scientific Research. The JCMT Transient Survey project codes are M16AL001 and M20AL007.

This research used the facilities of the Canadian Astronomy Data Centre operated by NRC Canada with the support of the Canadian Space Agency. D.J.\ is supported by NRC Canada and by an NSERC Discovery Grant. G.J.H.\ is supported by general grant 12173003 awarded by the National Science Foundation of China. H.S.\ is supported by Institute of Astronomy and Astrophysics, Academia Sinica, and by grants from the Ministry of Science and Technology (MoST) in Taiwan through 109-2112-M-001-028- and 110-2112-M-001-019-.

\bibliographystyle{aasjournal}
\bibliography{TrueTransients}

\input{table1}
\input{table2}

\end{document}

%% file: table1.tex
%
%
\begin{deluxetable*}{lccccccc}[t]
\tablecaption{\label{tab:regions} Summary of JCMT Transient Survey Regions at 850~\micron}
\tablehead{
\colhead{Region} & 
\colhead{R.A.} & 
\colhead{Dec.} & 
\colhead{Epochs} &
\colhead{First} & 
\colhead{Latest} &
\colhead{Epoch} & 
\colhead{Mean Brightness} \\
\colhead{} & 
\colhead{} & 
\colhead{} & 
\colhead{} &
\colhead{} & 
\colhead{} &
\colhead{Uncertainty} & 
\colhead{Uncertainty} \\
\colhead{} & 
\colhead{J$2000$} & 
\colhead{(J2000)} & 
\colhead{} &
\colhead{Epoch} &
\colhead{Epoch} &
\colhead{mJy/beam} & 
\colhead{mJy/beam} }
\startdata
IC\,$348$      & 03:44:18 & $+$31:16:52 & 44 & 22/12/2015 & 03/03/2021 & 11 -- 16 & 2.0 \\
NGC\,$1333$    & 03:28:54 & $+$31:16:52 & 45 & 22/12/2015 & 03/03/2021 & 11 -- 13 & 1.8  \\
NGC\,$2024$    & 05:41:41 & $-$01:53:51 & 43 & 26/12/2015 & 06/04/2021 & 11 -- 12 & 1.9  \\
NGC\,$2068$    & 05:46:13 & $-$00:06:05 & 52 & 26/12/2015 & 17/04/2021 & 11 -- 14 & 1.6 \\
OMC\,$2/3$     & 05:35:33 & $-$05:00:32 & 42 & 26/12/2015 & 06/04/2021 & 11 -- 13 & 1.8  \\
Ophiuchus     & 16:27:05 & $-$24:32:37 & 34 & 15/01/2016 & 15/05/2021 & 11 -- 14 & 2.7  \\
Serpens\,Main  & 18:29:49 & $+$01:15:20 & 67 & 02/02/2016 & 09/06/2021 & 11 -- 12 & 1.1 \\
Serpens\,South & 18:30:02 & $-$02:02:48 & 52 & 02/02/2016 & 02/06/2021 & 11 -- 13 & 1.5
\enddata
\end{deluxetable*}     

%% file: table2.tex
%
%
\begin{deluxetable*}{lcrcc}[t]
\tablecaption{\label{tab:stats} JCMT Transient Survey Regions Statistics}
\tablehead{
\colhead{Region} & 
\colhead{Faint} & 
\colhead{YSOs} & 
\colhead{YSO} &
\colhead{Candidate}\\
\colhead{} & 
\colhead{Coverage} & 
\colhead{} & 
\colhead{Coverage} &
\colhead{Detections}\\
\colhead{} & 
\colhead{[\%]} & 
\colhead{} & 
\colhead{[\%]} &
\colhead{[$\sigma$]}
}
\startdata
IC\,$348$      &  99.7& 133& 0.8& -- \\
NGC\,$1333$    &  98.7&  80& 0.5& 3.3\\
NGC\,$2024$    &  98.0& 193& 1.1& -- \\
NGC\,$2068$    &  98.9&  97& 0.6& 4.4\\
OMC\,$2/3$     &  95.9& 358& 2.2& 25.9, 4.0, 3.7\\
Ophiuchus     &  98.5& 109& 0.7& -- \\
Serpens\,Main  &  99.0&  56& 0.3& 4.0, 3.0\\
Serpens\,South &  98.4& 217& 1.3& --
\enddata
\end{deluxetable*} 